\documentstyle[amssymb,aps,twocolumn,10pt]{revtex}

\begin{document}
\title{Role of Wood anomalies in optical properties of thin metallic films with a
bidimensional array of subwavelength holes}
\author{Micha\"{e}l SARRAZIN$^{*}$, Jean-Pol VIGNERON$^{*}$, Jean-Marie VIGOUREUX$%
^{\dagger }$}
\address{$^{*}$Laboratoire de Physique du Solide\\
Facult\'{e}s Universitaires Notre-Dame de la Paix\\
Rue de Bruxelles 61, B-5000 Namur, Belgium\\
$^{\dagger }$Laboratoire de Physique Mol\'{e}culaire, UMR-CNRS 6624\\
Universit\'{e} de Franche-Comt\'{e}\\
F-25030 Besan\c {c}on Cedex, France}
\maketitle

\begin{abstract}
Recents works deal with the optical transmission on arrays of subwavelength
holes in a metallic layer deposited on a dielectric substrate. Making the
system as realistic as possible, we perform simulations to enlighten the
experimental data. This paper proposes an investigation of the optical
properties related to the transmission of such devices. Numerical
simulations give theoretical results in good agreement with experiment and
we observe that the transmission and reflection behaviour correspond to
Fano's profile correlated with resonant response of the eigen modes coupled
with nonhomogeneous diffraction orders. We thus conclude that the
transmission properties observed could conceivably be explained as resulting
from resonant Wood's anomalies.
\end{abstract}

\pacs{78.20.-e, 42.79.Dj, 42.25.Bs, 73.20.Mf}

\section{Introduction}

Recent papers deal with optical experiments and simulations with various
metallic gratings constituted of a thin metallic layer deposited on a
dielectric substrate [1-18]. Such materials are typically one- or
two-dimensional photonic crystals with a finite spatial extension in the
direction perpendicular to the plane where the permittivity is periodic.

One-dimensional gratings have been widelely studied in particular on account
of interesting effects knows as Wood's anomalies [19-30]. As shown by A.
Hessel and A.A. Oliner [19] this effect take two distinct forms. One occurs
in diffraction gratings at Rayleigh's wavelengths if a diffracted order
becomes tangent to the plane of the grating. The diffracted beam intensity
increases just before the diffracted order vanishes. The other is related to
a resonance effect [19]. Such resonances come from coupling between the
non-homogeneous diffraction orders and the eigenmodes of the grating. Both
types of anomalies may occur separately and independently, or appear
together. M. Nevi\`{e}re and D. Maystre [20,21] presented a wide study of
the causes of Wood's anomalies. In addition to the Rayleigh's wavelengths
they discovered two another possible origins of such anomalies. One, called
''plasmons anomalies'', occurs when the surface plasmons of a metallic
grating are excited. The other appears when a dielectric coating is
deposited on a metallic grating and corresponds to guided modes resonances
in the dielectric layer. In fact, both anomalies correspond to differents
cases of the resonance effect report by A. Hessel and A.A. Oliner [19]. As
shown by A. Hessel and A.A. Oliner [19], depending on the type of periodic
structure the two kind of anomalies (i.e. Rayleigh's anomalies or resonant
anomalies) may occur separately or are almost superimposed. At last, we note
that these concepts have been first suggested by V. U Fano [22].

In this paper we perform simulations to examine the behaviour of the optical
properties of a device which consists of arrays of subwavelength cylindrical
holes in a chromium layer deposited on a quartz substrate (Fig. 1). The
values of permittivity being those obtained from experiments [31]. We
present the key role of Rayleigh's wavelength and eigenmodes resonances in
the behaviour of the zeroth order reflexion and the transmission.

Our numerical study rests on the following method. Taking into account the
periodicity of the device, the permittivity is first described by a Fourier
series. Then, the electromagnetic field is described by Bloch's waves which
can too be described by a Fourier series. In this context, Maxwell's
equations take the form of a matricial first order differential equation
along to the $z$ axis perpendicular to the $x$ and $y$ axis where the
permittivity is periodic [32,33]. The heart of the method is to solve this
equation. One approach deals with the propagation of the solution step by
step by using the scattering matrix formalism. More explicitly, we
numerically divide the grating along to the $z$ axis into many thick layers
for which we calculate the scattering matrix. The whole scattering matrix of
the system is obtain by using a special combination law applied step by step
to each S matrices along to the $z$ axis. Indeed, it is well know that S
matrices and their combinations are much better conditionned than transfert
matrices [33]. Note that our algorithm has been compared with accuracy with
others method such as FDTD or KKR [34]. In the present work the convergence
is obtain from two harmonics only, i.e. for 25 vectors of the reciprocal
lattice. Further more, here there is no convergence problem associated with
discontinuities such that we need to use Li's method [35,36].

In the following, for a square grating of parameter $a$, note that, $%
\overrightarrow{g}=\frac{2\pi }a\left( i\overrightarrow{e}_x+j%
\overrightarrow{e}_y\right) $, such that the couple of integers $(i,$ $j)$
denotes the corresponding vector of the reciprocal lattice, i.e. diffraction
order.

Reflected and transmitted amplitudes are linked to the incident field by the
use of the $S$ scattering\ matrix which is calculated by solving Maxwell's
equation using a Fourier series [32].\ Let us define $F_{scat}$ as the
scattered field, and $F_{in}$ as the incident field, such that 
\begin{equation}
F_{scat}=\left[ 
\begin{array}{c}
\overline{N}_d^{+} \\ 
\overline{X}_d^{+} \\ 
\overline{N}_v^{-} \\ 
\overline{X}_v^{-}
\end{array}
\right] \text{, \ \ \ \ \ }F_{in}=\left[ 
\begin{array}{c}
\overline{N}_v^{+} \\ 
\overline{X}_v^{+} \\ 
\overline{N}_d^{-} \\ 
\overline{X}_d^{-}
\end{array}
\right]  \label{incscat}
\end{equation}
where $\overline{A}$ is a vector containing all the component $A_{%
\overrightarrow{g}}$. The subscipts $v$ and$\ d$ are written for ''vacuum''
and ''dielectric substrate'' respectively, and the superscripts $+$ and $-$
denote the positive and negative direction along the $z$ axis for the field
propagation. For each vector $\overrightarrow{g}$ of the reciprocal lattice, 
$N_{v\overrightarrow{g}}^{-}$ and $X_{v\overrightarrow{g}}^{-}$ are the $s$
and $p$ amplitudes of the reflected field, respectively, and $N_{d%
\overrightarrow{g}}^{+}$ and $X_{d\overrightarrow{g}}^{+}$ , that of the
transmitted field in the device. On the same way, $N_{v\overrightarrow{0}%
}^{+}$ and $X_{v\overrightarrow{0}}^{+}$ define the $s$ and $p$
polarizations amplitudes of the incident field, respectively. Then, $%
F_{scat} $\ is connected to $F_{in}$ {\it via} the scattering matrix such
as: 
\begin{equation}
S(\lambda )F_{in}(\lambda )=F_{scat}(\lambda )  \label{S}
\end{equation}

Then, the flux $J$ of the Poynting vector through a unit cell area $\sigma $%
, for a incident homogeneous plane wave is given by: 
\begin{equation}
J_v^{+}=\frac \sigma {2\mu _0\omega }k_{v\overrightarrow{0}z}\left[ \left|
N_{v\overrightarrow{0}}^{+}\right| ^2+\left| X_{v\overrightarrow{0}%
}^{+}\right| ^2\right]  \label{Jv+}
\end{equation}
\begin{eqnarray}
J_d^{+} &=&\frac \sigma {2\mu _0\omega }\sum_{\overrightarrow{g}}k_{d%
\overrightarrow{g}z}\left[ \left| N_{d\overrightarrow{g}}^{+}\right|
^2+\left| X_{d\overrightarrow{g}}^{+}\right| ^2\right]  \label{Jd+} \\
&&\times \Theta (\varepsilon _d(\omega )\frac{\omega ^2}{c^2}-\left| 
\overrightarrow{k}_{//}+\overrightarrow{g}\right| ^2)  \nonumber
\end{eqnarray}
\begin{eqnarray}
J_v^{-} &=&-\frac \sigma {2\mu _0\omega }\sum_{\overrightarrow{g}}k_{v%
\overrightarrow{g}z}\left[ \left| N_{v\overrightarrow{g}}^{-}\right|
^2+\left| X_{v\overrightarrow{g}}^{-}\right| ^2\right]  \label{Jv-} \\
&&\times \Theta (\frac{\omega ^2}{c^2}-\left| \overrightarrow{k}_{//}+%
\overrightarrow{g}\right| ^2)  \nonumber
\end{eqnarray}
where the electromagnetic field has been written as a Fourier series [32]. $%
\Theta (x)$ is the Heaviside function which gives $0$ for $x<0$ and $+1$ for 
$x>0$. $\overrightarrow{k}_{//}$ and $\omega $ are the wave vector component
parallel to the surface, and the pulsation of an incident plane wave on the
system, respectively. We also define 
\begin{equation}
k_{u,\vec{g},z}=(\varepsilon _u(\frac \omega c)^2-|\vec{k}_{//}+\vec{g}%
|^2)^{1/2}  \label{kgz}
\end{equation}
where $\varepsilon _u$\ represents either the permittivity of the vacuum ($%
\varepsilon _v$), or of the dielectric substrate ($\varepsilon _d$). We note
that if $k_{u,\vec{g},z}\ $becomes imaginary then diffraction orders becomes
non-homogeneous. The wavelength ($\lambda =\frac{2\pi c}\omega )$ values
such that $k_{u,\vec{g},z}=0$ are called Rayleigh's wavelength.

We define the zeroth order transmission and reflection as, 
\begin{equation}
T_{(0)}=\frac \sigma {2\mu _0\omega J_v^{+}}k_{d\overrightarrow{0}z}\left[
\left| N_{d,\overrightarrow{0}}^{+}\right| ^2+\left| X_{d,\overrightarrow{0}%
}^{+}\right| ^2\right]  \label{T0}
\end{equation}
and 
\begin{equation}
R_{(0)}=-\frac \sigma {2\mu _0\omega J_v^{+}}k_{v\overrightarrow{0}z}\left[
\left| N_{v\overrightarrow{0}}^{-}\right| ^2+\left| X_{v\overrightarrow{0}%
}^{-}\right| ^2\right]  \label{R0}
\end{equation}
Moreover, numerical computation of the poles of $S(\lambda )$ is important
in order to study the eigenmodes of the structure$.$ Let us write eq.8 as 
\begin{equation}
S^{-1}(\lambda )F_{scat}(\lambda )=F_{in}(\lambda )  \label{S-1}
\end{equation}
In this way, the eingenmodes of the structure are solution of eq. 9 in the
case $F_{in}(\lambda )=0$, i.e. 
\begin{equation}
S^{-1}(\lambda )F_{scat}(\lambda )=0  \label{eigen}
\end{equation}
This is a typical homogeneous problem, well know in the theory of gratings
[20,21,36,37]. Complex wavelengths $\lambda _\eta =\lambda _\eta ^R+i\lambda
_\eta ^I$, for which eq.(10) has non-trivial solutions, are the poles of $%
\det (S(\lambda ))$ as we have 
\begin{equation}
\det (S^{-1}(\lambda _\eta ))=0  \label{S-10}
\end{equation}
In this way, if we extract the singular part of $S$ corresponding to the
eigenmodes of the structure, we can write $S$ on a analytical form as
[20,21,37,38] 
\begin{equation}
S(\lambda )=\sum_\eta \frac{R_\eta }{\lambda -\lambda _\eta }+S_h(\lambda )
\label{laurent}
\end{equation}
This is a generalized Laurent series, where $R_\eta $ are the residues
associated which each poles $\lambda _\eta $. $S_h(\lambda )$ is the
holomorphic part of $S$ which corresponds to purely non-resonant processes.

Thus, assuming that $f(\lambda )$ is the $m^{th}$ component of $%
F_{scatt}(\lambda )$, we have, for the expression of $f(\lambda )$ in the
neighboorhood of one pole $\lambda _\eta $ [20,21,37,38] 
\begin{equation}
f(\lambda )=\frac{r_\eta }{\lambda -\lambda _\eta }+s(\lambda )
\end{equation}
where $r_\eta =$ $\left[ R_\eta F_{in}\right] _m$ and $s(\lambda )=\left[
S_h(\lambda )F_{in}\right] _m$.

\section{Results}

The calculated transmission against the wavelength of the incident wave on
the surface is shown in Fig.\ 2 for the zeroth diffraction order, for light
incidence normal to the surface and electric field polarized parallel to $x$
axis. The diameter of holes ($d=500nm)$ and the thickness of the film ($%
h=100nm$) have been chosen according to the experimental conditions [1,2].
The solid, and dashed lines\ represent the transmission for a square grating
of parameter $a=1$ and $1.2\mu m$, respectively, whereas the dotted line
corresponds to the transmission for a similar system without holes. In Fig.\
2, it is shown that the transmission increases with the wavelength, and that
it is characterized by sudden changes in the transmission marked 1 to 4 on
the figure. If wavelength 1,2 and 4 correspond to minima, the wavelength 3
is nevertheless not explicitely a minimum, as we will explain it later.
These values are shifted toward larger wavelengths when the grating size
increases, and the minima disappear when considering a system without hole.
Note that these results qualitatively agree with the experimental data of
Ebbesen et al. [1,2]. Values of wavelength marked 1 to 4 are given in the
first column of table 1. In the second column we give the values of the
positions of maxima marked A to C on the figure.

In Fig. 3 we give the calculated reflection as a function of the wavelength
of the incident wave on the surface for the zeroth diffraction order, for
both gratings and for the system without holes. The reflection curves are
characterized by maxima (numbered 1 to 4) which correspond to the minima
calculated in the transmission curves.\ On the same way, the location of
these maxima are shifted towards larger wavelengths when the grating size
increases, and they disappear when the surface is uniform. Then, it appears
that the sudden decreases in transmission is correlated to an increased
reflection.\ Moreover, the positions of the correlated maxima and minima are
calculated at wavelengths which seems to correspond to Rayleigh's
wavelengths as shown in the first column of Table 2. We have reported the
positions of the maxima of transmission, marked A to C, on the Fig. 3. We
note that the maxima in transmission are not correlated with specifics
values of reflection.

In Fig. 4 we give the calculated absorption against the wavelength of the
incident wave on the surface, for the zeroth diffraction order. The solid
line denotes the absorption for the square grating of parameter $a=1\mu m$,
the dashed line denotes the absorption for the square grating of parameter $%
a=1,2\mu m$, and the dotted line denotes the absorption of a similar system
without holes. We have reported the positions of minima in Fig. 2, numbered
1 to 4, and the positions of points A to C which denote the maxima. These
peaks are found at longer wavelengths when the grating size increases, and
they disappear when the surface is uniform. Thus it appears that the sudden
decrease in reflectance is caused by a combination of increased reflectance
and increased loss due to surface roughness.

Previous work [1-18] have identified the convex regions in transmittance,
i.e., those regions between the local minima, as regions where plasmons
exist. If this were indeed the case, then we would expect to observe local
maxima in the loss of energy. However, if we compare figure 2 with figure 4,
we see that the convex regions in figure 2 are not matched by increased loss
in figure 4, nevertheless the maxima of absorption seems to correspond to
the minima of transmission.

On the basis of these results, we investigate the role of Wood's anomalies
in the physical interpretation of our simulations. In this way, we emphazise
the existence of eigenmodes and their role {\it via} resonant coupling with
the electromagnetic field.

First, we are stuying poles and resonances of the grating. As explained in
the introduction the existence of eigenmodes is linked to the existence of
poles of the scattering matrix. If we make the assumption that the role of
purely non-resonant process is negligible, i.e. $s(\lambda )\sim 0$, then
eq. (13) can be approximated by the following expression [20,21,37,38] in
the vicinity of one pole $\lambda _\eta ^R+i\lambda _\eta ^I$%
\begin{equation}
\left| F_{scat}(\lambda )\right| \thicksim \frac{\left| r_\eta \right| }{%
\sqrt{\left( \lambda -\lambda _\eta ^R\right) ^2+\lambda _\eta ^{I\ \ 2}}}
\end{equation}
which gives a typical resonance curve where the wavelength of resonance $%
\lambda _r$ is equal to $\lambda _\eta ^R$, and where the width $\Gamma $ at 
$\frac 1{\sqrt{2}}\left| F_{scat}(\lambda _r)\right| $ is equal to $2\lambda
_\eta ^I$. Before searching for typical resonance in the behaviour of
diffraction orders we check the existence of poles of the $S$ matrix.

In the third column in Table 2, we give the poles $\lambda _\eta =\lambda
_\eta ^R+i\lambda _\eta ^I$ of the $S$ matrix computed numerically. We keep
only the values whose real part is close to the values (1) to (4) in Fig. 2
and Fig. 3. This result suggests the possibility of resonant processes. In
order to investigate such assumption, we have studied the behavior of the
intensity of some specific diffraction orders on the vacuum/metal and
substrate/metal interfaces.\ More precisely, we have considered the
diffraction orders corresponding to the Rayleigh's wavelengths connected to
the positions of the minima obtained in the transmission curves. We compare
the results with the transmission and reflection curves.

In Fig. 5, curve (a) shows the modulus of the electromagnetic field of the
orders $(\pm 1,0)$ at the substrate/metal interface, as a function of the
wavelength. The same is true for the curve (b) but the interface is now
vacuum/metal. Curve (c) shows the reflected $(0,0)$ order. One notices the
presence of localized peaks in curves (a) and (b). Simulations allow one to
check that orders $(\pm 1,0)$ have only $p$ polarization. These peaks
coincide with the minima of the curve of transmission of Fig. 2. Since these
peaks correspond to orders with $p$ polarization, they are probably
resonances of the structure. To confirm this, we evaluate the poles by
measuring the wavelength of resonance $\lambda _r$ (which is equal to $%
\lambda _\eta ^R$), and the width $\Gamma $ at $\frac 1{\sqrt{2}}\left|
F_{scat}(\lambda _r)\right| $ (which is equal to $2\lambda _\eta ^I$). We
obtain results given in the fourth column in Table 2. One can easilly
compare these results with those of the third column in table 2. This
confirms the resonant characteristic of the diffraction orders $(\pm 1,0)$\
at the metal/vacuum and metal/substrate interfaces (A. Hessel and A.A.
Oliner has called such diffraction orders ''resonant diffraction orders''
[19]). Note that the orders $(\pm 1,0)$ at the vacuum/metal interface and $%
(\pm 1,\pm 1)$ at the substrate/metal interface has poles with closer real
part. This means that both modes are almost degenerated with the consequence
that both modes effects can't be clearly distinguished particularly for the
transmission. So, the wavelength (3) does not seem to provide a minimum as
clearly as the wavelength (2).

Fig. 6 shows the behavior of the amplitude modulus of the diffraction orders 
$(0,\pm 1)$ for the vacuum/metal (curve (a)) and for the substrate/metal
(curve (b)) interfaces, respectively, as a function of the wavelength. Curve
(c) corresponds to the order $(0,0)$ in transmission. All these orders exist
only with a polarization $s.$ One notices that the minima of these curves
are correlated with the peaks of resonances. On the other hand, we know that
orders with $s$ polarization cannot present resonances. From this point of
view, the minima of the curve of transmission of the Fig. 2 are correlated
with the resonances, while the behavior of the convex parts of the curves of
transmission can be interpreted according to the profile of the orders of
polarization $p$.

Let us now turn to Wood's anomalies. We consider the case where purely
non-resonant process can't be totaly neglected such that we suppose $%
s(\lambda )\sim s_0$. Thus, it is easy to show that eq.(13) can be written
as [19,22] 
\begin{equation}
\left| F_{scat}(\lambda )\right| ^2=\frac{\left( \lambda -\lambda
_z^R\right) ^2+\lambda _z^{I\ \ 2}}{\left( \lambda -\lambda _\eta ^R\right)
^2+\lambda _\eta ^{I\ \ 2}}\left| s_0\right| ^2
\end{equation}
with
\begin{equation}
\lambda _z^R=\lambda _\eta ^R-\nu ^R\text{ \ \ \ and \ \ }\lambda
_z^I=\lambda _\eta ^I-\nu ^I
\end{equation}
where
\begin{equation}
\nu =\frac{r_\eta }s
\end{equation}
Coefficient $\nu $ shows the significance of resonant effect compared with
purely non-resonant effects. $\lambda _z=\lambda _z^R+i\lambda _z^I${\it \ }%
corresponds to the zero of eq.{\it \ }(13) and (15). Eq. (15) corresponds to
the profiles of Fig. 7. This last expression takes into account the
interferences between resonant and purely non-resonant processes. In this
way, the profiles which correspond to the eq. (14), i.e. a purely resonant
process, tend to become asymmetric. As shown in Fig. 7, dashed curve shows a
typically resonant process like those described by eq.(14). On the other
hand, solid and dash-dotted curves show a typical behaviour where a minimum
is followed by a maximum, and vice versa assuming the values of $\nu $.
These profiles tend to $\left| s_0\right| ^2$ when $\lambda $ tends to $\pm
\infty $. We note that these properties, which result from the interference
of resonant and non-resonant processes, are similare to those described by
A.A. Hessel, A. Oliner [19] and V.U. Fano [22]. For this reason the profiles
like those described on Fig. 7 are often called ''Fano's profiles''.

In order to refine the interpretation of our results, we represent on Fig.\
8 the three curves (transmission, reflection and resonant diffraction order)
on a more restricted domain of wavelength in the range $1300-1900$ $nm$. In
this range, since Rayleigh's wavelength is associated to the resonant
diffraction order $(1,0)$ for the metal/substrate interface, we represent
the amplitude of this order only. The solid line denotes the transmission,
the dashed line denotes the reflection and the dash-dotted line denotes the
amplitude of the resonant diffraction order. We also indicate the position
of the corresponding Rayleigh wavelength, as well as that of the maximum of
resonance (vertical dotted lines). One labels $(a)$ the maximum of the
transmission, $(b)$ the minimum of the reflection and $(c)$ the maximum of
the transmission.

One notices that the maximum of resonance does not strictly coincide with
the maximum of reflection and the minimum of transmission. Also, one notices
that the maximum of reflection does not coincides with the minimum of
transmission. On the other hand Rayleigh's wavelength seems well to
correspond with the minimum of transmission. We notice that the diffraction
order is homogeneous for wavelengths lower than Rayleigh's wavelength. For
this reason, the resonance peak cannot be observed for wavelengths lower
than Rayleigh's wavelengths. So, if one intends to take away the position of
resonance of the value of Rayleigh's wavelength, one can make it {\it a
priori} only in the direction of increasing wavelengths. Should the opposite
occur, the position of the resonance peak tends towards Rayleigh's value.

As in Fig.\ 8, we represent on Fig.\ 9\ the three curves (transmission,
reflection and resonant diffraction order) for the same physical parameters.
However, whereas in the previous case the value of the permittivity of the
metal film was that of chromium [31], we use now the value equal to $-25+i1$
which does not depend of the wavelength. Such value of the permittivity does
not correspond to an existing material. We just choose this permittivity
value for the metal such that we select a peak of resonance farther from
Rayleigh's wavelength than in the previous case. The choice of this value
only comes from the research of the compromise between the position from the
peak of resonance and its width so as to illustrate our matter clearly. As
in Fig.\ 8, $(a)$ is the maximum of the transmission, $(b)$ the minimum of
the reflection and $(c)$ the maximum of the transmission. One names $(d)$
the minimum of the transmission.

This time, one notices in a clear way the absence of coincidence between the
peak of resonance and the minima (respectively the maxima) of reflection
(respectively of transmission). Contrary to what is generally assumed
[1-16], one sees that the nonresonant Wood's anomalies connected to
Rayleigh's wavelengths are not the cause of the minima of transmission. They
simply correspond to a discontinuity of each of the three curves. It is
particularly important to note that the profiles of the transmission and the
reflection correspond to Fano's profiles as discussed below. One can
interpret the behavior of these spectra in term of resonant Wood's anomalies
in the sense described by V. U. Fano [22] and by A. Hessel and A.A. Oliner
[19].

\section{Discussion}

In order to understand the physical mechanisms responsible for the behavior
observed on Fig. 8 and Fig. 9, we have represented in Fig. 10 the
corresponding involved processes. On Fig. 10, circles $A$ and $B$ represent
diffracting elements (e.g. holes). So, an incident homogeneous wave $\left(
i\right) $ diffracts in $A$ and generates a nonhomogeneous resonant
diffraction order $\left( e\right) $ (e.g. $\left( 1,0\right) $). Such order
is coupled with a eigenmode which is characterized by a complex wavelength $%
\lambda _\eta $. It becomes possible to excite this eigenmode which leads to
a feedback reaction on the order $\left( e\right) $. This process is related
to the resonant term.

The diffraction order $\left( e\right) $\ diffracts in $B$ and generates a
contribution to the homogenous zero diffraction order $\left( 0,0\right) $.
Thus, one can ideally expect to observe a resonant profile, i.e. lorentzian
like, for the homogenous zero diffraction order $\left( 0,0\right) $ which
appears in $B$. Nevertheless, it is necessary to account for nonresonant
diffraction processes related to the holomorphic term. So, incident wave $%
(i) $, here represented in $B$, generates a homogeneous zero order. Then,
one takes into account the interference of two rates, resonant and non
resonant contribution to zero order. The resulting lineshapes are typically
the Fano's profiles which correspond to resonant process where one takes
into account nonresonant effects. One notes that a maximum in transmission
does not necessary correspond to the maximum of resonance of a diffraction
order. It is exactly the process observed on Fig. 9 where the resonance is
associated with the diffraction order $\left( 1,0\right) $. So, the Fano's
profiles of the reflection and the transmission, result from the
superimposing of resonant and nonresonant contribution to the zero
diffraction order.

If one refers to Fig. 8, the concrete case of the chromium, the resonance is
closer to Rayleigh's wavelength than in the case of the Fig. 9. In another
hand the positions of the maximum and the minimum of a Fano's profile are
determined by the resonance position. More precisely, if the resonance is
shifted in a given direction, the maximum and the minimum of the Fano's
profile tends to be shifted in the same way. Consequently, in the present
case, the maximum and the minimum of the asymmetric Fano's profile is
shifted toward Rayleigh's wavelength in the same way as the resonant
response. In Fig. 8, in the case of the transmission, minimum (d) is not of
the same kind of the minimum (d) in Fig. 9. This is not a true minima of the
Fano's profile. All occurs like if the minimum of the Fano's profile
disappears behind the Rayleigh's wavelength towards low wavelength. In other
words, the minimum (d) in Fig. 8 comes from the cut off and the
discontinuity introduce between the minimum and the maximum of the Fano's
profile at the Rayleigh's wavelength. On the other hand, note that maximum
(a) of the transmission and maximum (c) of the reflection just localized
rests after Rayleigh's wavelength. For the reflection, minima (b) tends to
be shifted towards low wavelength.

Previous works [1-16] have identified the convex regions in transmission,
i.e., the regions between the minima, as regions where plasmons exist. The
present study tends to qualify this hypothesis, since it shows that the
experimental results can be described in terms of Wood's anomalies. Indeed,
as shown by V.U. Fano [22], A. Hessel and A.A. Oliner [19], for one
dimensionnal gratings, Wood's anomalies can be treated in terms of
eigenmodes grating excitation. In this context, these authors demonstrated
the asymmetric behavior of the intensities of the homogeneous diffraction
orders according to the wavelength. One can conclude that the results of
T.W. Ebbesen's experiments correspond to the observation of resonant Wood's
anomalies.

Here, as we use metal in our device, it seems natural to assume that these
resonances are surface plasmons resonances. Nevertheless, it is important to
note that our analysis don't make any hypothesis on the origin of the
eigenmodes. This involve that it could be possible to obtain transmission
curves similar to those for metals, by substituted the surface plasmons by
polaritons or guided modes. This work is in progress.

\section{Conclusion}

Using a system similar to that used in recent papers [1,2], we have shown
that numerical simulations give theoretical results in good qualitative
agreement with experiments. Previous authors have suggested that the results
are due to the presence of the metallic layer, such that the surface
plasmons could give rise to transmission curves of these characteristics. We
have performed simulations using the same geometry, and we have observed
that the transmission and reflection behaviour correspond to Fano's profiles
correlated with resonant response of the eigenmodes coupled with
nonhomogeneous diffraction orders. We thus conclude that the transmission
properties observed could conceivably be explained as resulting from
resonant Wood's anomalies.

[1] T.W. Ebbesen, H.J. Lezec, H.F. Ghaemi, T. Thio, P.A. Wolff, Nature
(London) 391, 667 (1998)

[2] T. Thio, H.F. Ghaemi, H.J. Lezec, P.A. Wolff, T.W. Ebbesen, JOSA B, 16,
1743 (1999)

[3] H.F. Ghaemi, T. Thio, D.E. Grupp, T.W. Ebbesen, H.J. Lezec, Phys. Rev.
B, 58, 6779 (1998)

[4] U. Schr\"{o}ter, D. Heitmann, Phys. Rev. B, 58, 15419 (1998)

[5] D. E. Grupp, H.J. Lezec, T. Thio, T.W. Ebbesen, Adv. Mater, 11, 860
(1999)

[6] T.J. Kim, T. Thio, T.W. Ebbesen, D.E. Grupp, H.J. Lezec, Opt. Lett, 24,
\ 256 (1999)

[7] J.A. Porto, F.J. Garcia-Vidal, J.B. Pendry, Phys. Rev. Lett., 83, 2845
(1999)

[8] Y. M. Strelniker, D. J. Bergman, Phys. Rev. B, 59, 12763, (1999)

[9] S. Astilean, Ph. Lalanne, M. Palamaru, Optics Comm. 175 (2000) 265-273

[10] D.E. Grupp, H.J. Lezec, T.W. Ebbesen, K.M. Pellerin, T. Thio, Applied
Physics Letters 77 (11) 1569 (2000)

[11] E. Popov, M. Nevi\`{e}re, S. Enoch, R. Reinisch, Phys. Rev. B, 62,
16100 (2000)

[12] W.-C. Tan, T.W. Preist, R.J. Sambles, Phys. Rev. B 62 (16) 11134 (2000)

[13] T. Thio, H.J. Lezec, T.W. Ebbesen, Physica B 279 (2000) 90-93

[14] A. Krishnan, T. Thio, T. J. Kim, H. J. Lezec, T. W. Ebbesen, P.A.
Wolff, J.\ Pendry, L. Martin-Moreno, F. J. Garcia-Vidal, Optics Comm., 200,
1-7 (2001)

[15] L. Martin-Moreno, F.J. Garcia-Vidal, H.J. Lezec, K.M. Pellerin, T.
Thio, J.B. Pendry, T.W. Ebbesen, Phys. Rev. Lett., 86, 1114 (2001)

[16] L. Salomon, F. Grillot, A.V. Zayats, F. de Fornel, Phys. Rev. Lett., 86
(6), 1110 (2001)

[17] M.M.J. Treacy, Appl. Phys. Lett., 75, 606, (1999)

[18] J.-M. Vigoureux, Optics Comm., 198, 4-6, 257 (2001)

[19] A. Hessel, A. A. Oliner, Applied Optics 4 (10) 1275 (1965)

[20] D. Maystre, M. Nevi\`{e}re, J. Optics, 8, 165 (1977)

[21] M. Nevi\`{e}re, D. Maystre, P. Vincent, J. Optics, 8, 231 (1977)

[22] V.U. Fano, Ann. Phys. 32, 393 (1938)

[23] R. H. Bjork, A. S. Karakashian, Y. Y. Teng, Phys. Rev. B, 9, 4, 1394
(1974)

[24] P.J. Bliek, L.C. Botten, R. Deleuil, R.C. Mc Phedran, D. Maystre, IEEE
Trans. Microwave Theory and Techniques, MTT-28 1119-1125 (1980)

[25] D. Deaglehole, Phys. Rev. Lett., 22, 14, 708 (1969)

[26] E. Popov, L. Tsonev, D. Maystre, Applied Optics 33 (22) 5214 (1994)

[27] Lord Rayleigh, Proc. Roy. Soc. (London) A79, 399 (1907)

[28] K. Utagawa, JOSA 69 (2) 333\ (1979)

[29] L. Wendler, T. Kraft, M. Hartung, A. Berger, A. Wixforth, M. Sundaram,
J. H. English, A. C. Gossard, Phys. Rev. B, 55, 4, 2303 (1997)

[30] R.W. Wood, Phys. Rev. 48, 928 (1935)

[31] D.W. Lynch, W.R. Hunter, in Handbook of Optical Constants of Solids II,
E.D. Palik, (Academic Press, Inc., 1991)

[32] J.P. Vigneron, F. Forati, D. Andr\'{e}, A. Castiaux, I. Derycke, A.
Dereux, Ultramicroscopy, 61, 21 (1995)

[33] J.B. Pendry, P.M. Bell, NATO\ ASI\ Series E Vol. 315 (1995)

[34] V. Lousse, K. Ohtaka, Private Communication (2001)

[35] L. Li, JOSA\ A 13\ (9) 1870 (1996)

[36] L. Li, JOSA A 14 2758 (1997)

[37] E. Centeno, D. Felbacq, Phys. Rev. B 62 (12) R7683 (2000)

[38] E. Centeno, D. Felbacq, Phys. Rev. B 62 (15) 10101 (2000)

[39] R. Petit, Electromagnetic Theory of Gratings, Topics in current
Physics, 22, Springer Verlag (1980)

\section{Captions}

FIG. 1 Diagrammatic view of the system under study. Transmission and
reflection are calculated for the zeroth order and at normal incidence as in
experiments.

FIG. 2 Percentage transmission of the incident wave against its wavelength
on the surface, for the zeroth diffraction order. The solid line denotes the
transmission for the square grating of parameter $a=1\mu m$, the dashed line
denotes the transmission for the square grating of parameter $a=1.2\mu m$,
and the dotted line denotes the transmission of a similar system without
holes. The points numbered 1 to 4 denote sudden changes in the transmission
whereas the points A to C denote the maxima.

FIG. 3 Reflection against the wavelength of the incident wave on the
surface, for the zeroth diffraction order. The solid line denotes the
reflection for the square grating of parameter $a=1\mu m$, the dashed line
denotes the reflection for the square grating of parameter $a=1,2\mu m$, and
the dotted line denotes the reflection of a similar system without holes. We
note that the minima in Fig. 2 are matched by peaks in the reflection (see
the previous figure), numbered 1 to 4. We have reported the points A to C
which denote the positions of maxima of the transmission.

FIG. 4 Absorption against the wavelength of the incident wave on the
surface, for the zeroth diffraction order. The solid line denotes the
absorption for the square grating of parameter $a=1\mu m$, the dashed line
denotes the absorption for the square grating of parameter $a=1,2\mu m$, and
the dotted line denotes the absorption of a similar system without holes. We
have reported the positions of minima in Fig. 2, numbered 1 to 4, and the
positions of points A to C which denote the maxima.

FIG. 5 Curve (a) shows the modulus of the electromagnetic field of the
orders $(\pm 1,0)$ at the substrate/metal interface, as a function of the
wavelength. Idem for the curve (b) but for the vacuum/metal interface. Curve
(c) shows the reflected $(0,0)$ order. One notices the presence of localized
peaks in curves (a) and (b). The amplitude of the incident field is equal to 
$1$ $V.m^{-1}$.

FIG. 6 Behavior of the amplitude modulus of the diffraction orders $(0,\pm
1) $ respectively for the interface vacuum/metal (curve (a)) and for the
interface substrate/metal (curve (b)) as a function of the wavelength. Curve
(c) corresponds to the order $(0,0)$ in transmission. All these orders exist
only with a polarization $s.$ The amplitude of the incident field is equal
to $1$ $V.m^{-1}$.

FIG. 7 Some examples of typical Fano's profiles.

FIG. 8 The set of three curves (transmission, reflection and resonant
diffraction order) on a more restricted domain of wavelength included
between $1300$ $nm$ and on $1900$ $nm$. In this domain Rayleigh's wavelength
is associated to diffraction order $(1,0)$ for the interface
metal/substrate. We also indicate the position of the wavelength of
corresponding Rayleigh as well as that of the maximum of resonance. One
names $(a)$ the maximum of the transmission, $(b)$ the minimum of the
reflection and $(c)$ the maximum of the reflection. Solid line :
transmission, dashed line : reflection, dash-dotted line : resonant
diffraction order. The amplitude of the incident field is equal to $1$ $%
V.m^{-1}$.

FIG. 9 Similar system than in Fig. 7 except for the value of the
permittivity of the metal film here equal to $-25+i1$. As in Fig. 7, $(a)$
is the maximum of the transmission, $(b)$ the minimum of the reflection and $%
(c)$ the maximum of the reflection. One names $(d)$ the minimum of the
transmission. The amplitude of the incident field is equal to $1$ $V.m^{-1}$.

FIG. 10 Diagrammatic representation of the processes responsible of the
behaviour of the transmission properties.

TABLE\ 1 : Positions of minima and maxima of transmission.

TABLE\ 2 : Comparison between Rayleigh's wavelengths (second column) of some
diffraction orders (first column) with the poles of the scattering matrix
computed numerically (third column) and evaluated by measuring the
wavelength of resonance $\lambda _r$, and the width $\Gamma $ of some
resonance curves (fourth column). $(v/m)$ and $(s/m)$ denote vacuum/metal
interface and substrate/metal interface respectively.

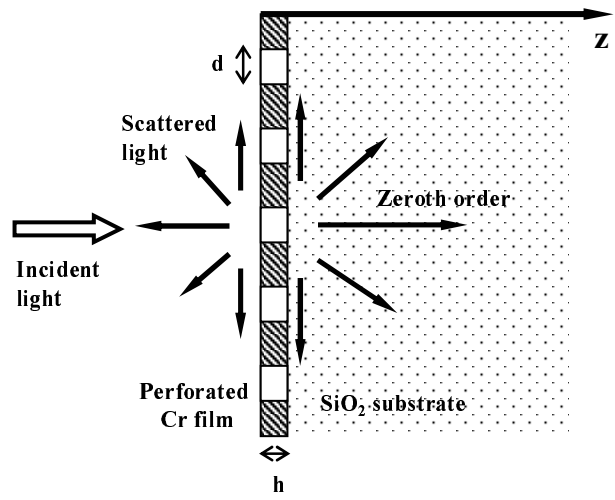
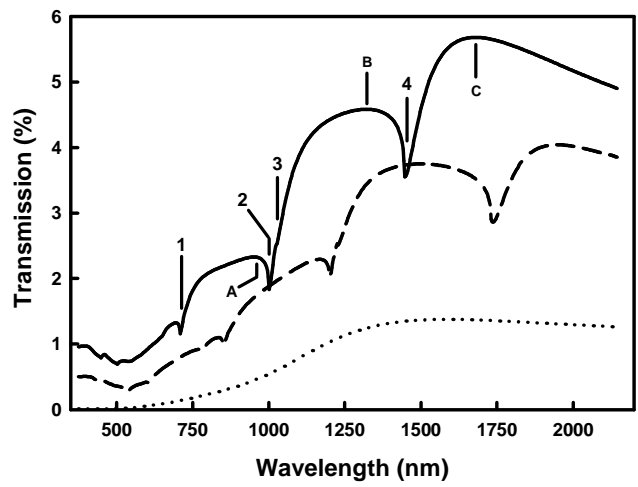
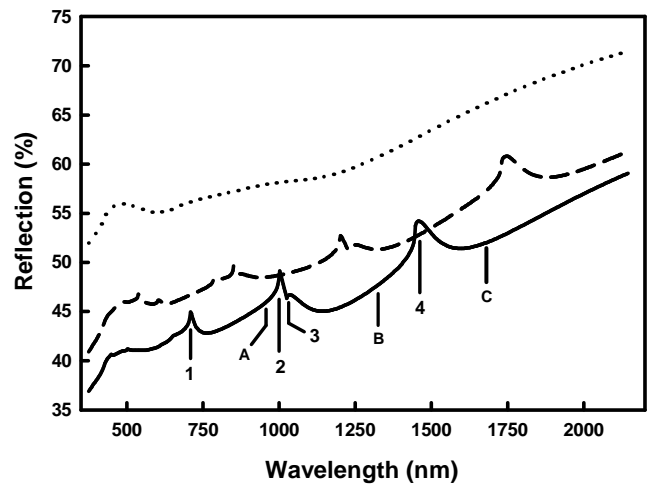
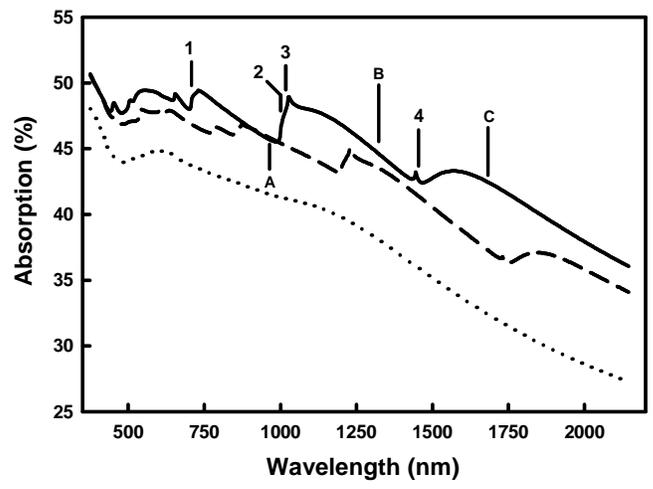
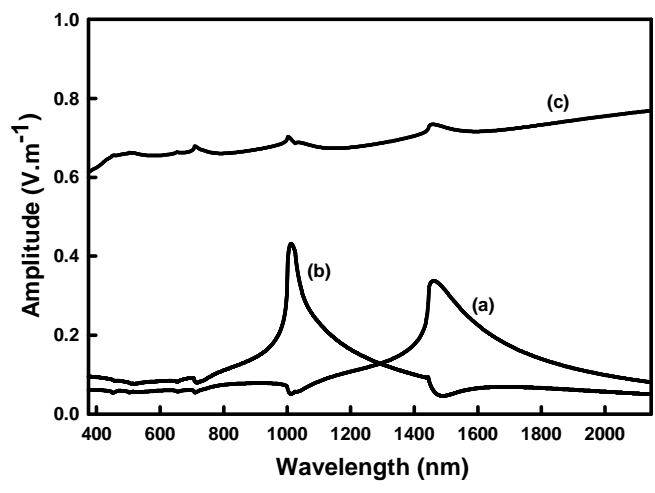
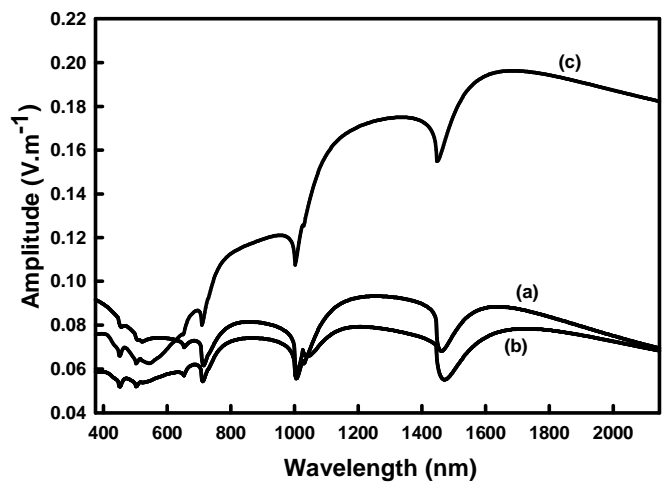
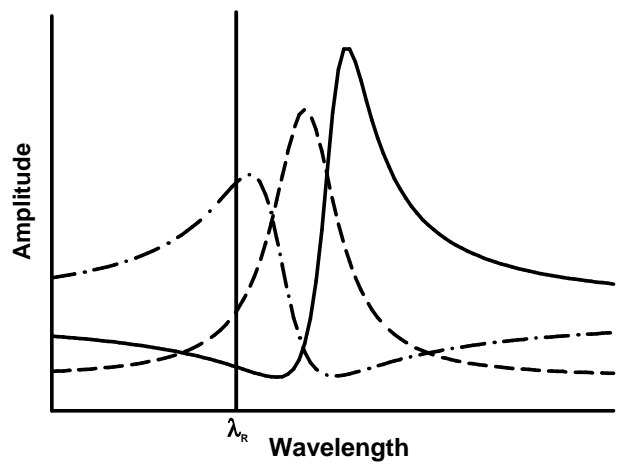
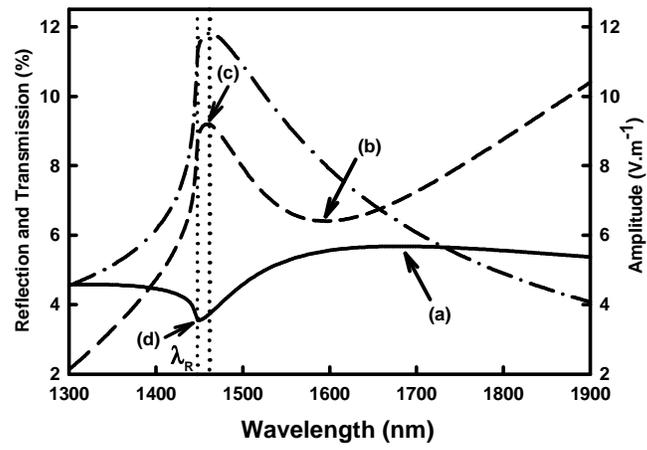
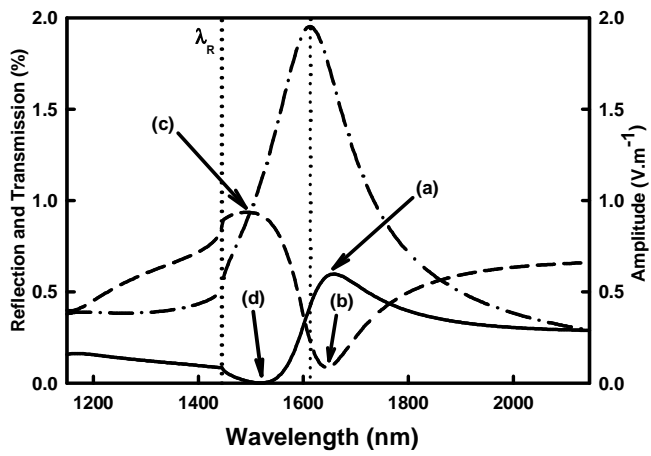
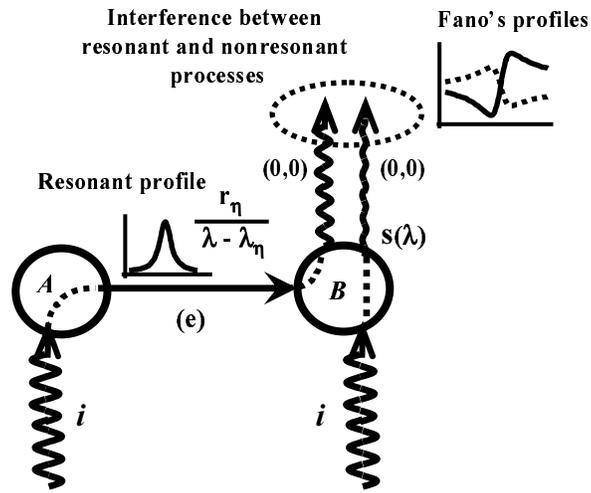
\end{document}